# Magnetization studies on superconducting $MgB_2$ - lower and upper critical fields and critical current density

Amish G. Joshi[a], C.G.S. Pillai[b], P. Raj[b] and S.K. Malik,[a]

[a]Tata Institute of Fundamental Research, Colaba, Mumbai, 400 005, India

[b]Novel Materials and Structural Chemistry Division, BARC, Trombay, Mumbai, 400 085, India

## Abstract

Magnetization studies have been carried out on superconducting $MgB_2$ ($T_c$=37K) in the temperature range of 2-50K and in magnetic fields up to 5 Tesla. From these measurements, values of the lower critical field $H_{c1}(0)$, upper critical field $H_{c2}(0)$ at zero K are estimated to be ~300 Oe and ~12.5 Tesla, respectively, which yield a value of ~26 for the Ginzburg-Landau parameter, κ. Magnetization hysteresis loops have been obtained at various temperatures from which the magnetization critical current density is estimated using Bean's critical state model.

## I Introduction

The recent discovery of superconductivity in $MgB_2$ with a superconducting transition temperature, $T_c$, close to 39K [1] has created a great deal of excitement in the area of condensed matter physics. This compound crystallizes in the hexagonal $AlB_2$ type structure, which consists of alternating layers of Mg atoms and boron honeycomb layers. The boron isotope effect studies [2] suggest that this compound is a conventional phonon mediated BCS superconductor. This is also supported by band structure calculations [3]. The origin of the high $T_c$ in this material seems to be associated with the light boron atoms whose phonon frequency spectrum plays an important role in enhancing the electron-phonon interaction. In this communication, we present the results of our magnetization studies on $MgB_2$ from which information about the lower and upper critical fields and the magnetization current density is obtained.

## II Experimental Details

The $MgB_2$ sample was prepared by heating together stoichiometric amounts of Mg and B. The Mg turnings and finely powdered boron were mixed and compacted in to a pellet. This pellet was wrapped in a tantalum foil and heated in a tubular furnace in a flowing atmosphere of commercially available 92% Ar, 8% $H_2$ gas mixture. The pellet was held at 600ºC for 2 hours, followed by 800ºC for 2 hours and finally at 950ºC for an hour before cooling to room temperature. The product was powdered, compacted and reheated at 950ºC for 2 hours and cooled to room temperature by switching off the furnace.

Powder X-ray diffraction pattern, obtained using $CuK_\alpha$ radiation, showed the sample to largely consist of desired $MgB_2$ phase. A small amount of MgO impurity phase was also found to be present - possibly due to the use of commercial Ar+$H_2$ gas mixture during heating. A Rietveld refinement of the x-ray pattern, based on the hexagonal $AlB_2$ type structure (space group *P6/mmm* #191), yielded lattice parameters *a*=3.0815(6)Å and *c*=3.5191(7)Å. Magnetization measurements on the powdered $MgB_2$ sample were carried out using a SQUID magnetometer (Quantum Design) in the temperature range 2-50K and in applied fields up to 5.5 Tesla. Some of the hysteresis loops were also obtained using a vibrating sample magnetometer (Oxford Instruments).

## III Results and Discussion

Figure 1 show a plot of magnetization versus temperature for $MgB_2$ in an applied filed of 50 Oe. The magnetization was measured on warming up the sample from 5K to 50K both in zero-filed-cooled (ZFC) and in field-cooled (FC) states. The onset of diamagnetism is observed at a temperature of about 37K, which is taken to be the superconducting transition temperature, $T_c$. This is somewhat lower than the $T_c$ of about 39K obtained by other workers [see, for instance, 1-2]. The lower $T_c$ in the present sample could be possibly due to the presence of small amount of MgO phase and the resulting off stoichiometry of the desired superconducting phase of $MgB_2$.

Figure 2 shows a plot of magnetization versus applied field at various temperatures in the ZFC state. The initial deviation of the magnetization from linearity, indicating the penetration of magnetic flux in the sample, is observed at fields of only few hundred gauss. To get better indication of the onset of deviation from linearity, we adopt a procedure used earlier [4]. First, a straight line fit to the very low field magnetization data is obtained. This straight line is extended to higher fields and the deviation of observed magnetization from this extrapolated linear curve is obtained and plotted as a function of field as shown in Fig. 3. The deviation in the magnetization from the straight-line fit is much more discernible in this plot from which values of $H_{c1}$ at various temperatures can be obtained. For instance, $H_{c1}$ value estimated at 5K is ~260 Oe. Inset in Fig. 3 shows the plot of $H_{c1}$ as a function of temperature. An interesting feature of the plot is the near linear increase of $H_{c1}$ with decreasing temperature. The extrapolated zero temperature value of $H_{c1}$ is ~300 Oe. Others [5-7] have reported similar values of $H_{c1}$.

Figure 4 shows a plot of ZFC magnetization versus temperature measured in various applied fields from which superconducting transition temperatures at various fields are obtained. These are used to create the $H_{c2}$ versus temperature plot shown as an inset in Fig. 4. The slope of the $dH_{c2}/dT$ curve between 1 Tesla and 5 Tesla field is found to be ~0.5T/K. To obtain the value of $H_{c2}$ at zero K, we use the WHH formula: $H_{c2}(0)=0.69T_c |dH_{c2}/dT|$ which yields $H_{c2}(0)=12.5$ Tesla, which is comparable to that obtained by other authors from resistive/magnetic measurements [6, 8, 9]. One may use the $H_{c1}$ and $H_{c2}$ data to obtain the value of Ginzburg-Landau parameter, $\kappa=\lambda/\xi$, by employing the relations $H_{c1}=(\phi_0/4\pi\lambda^2)\ln(\lambda/\xi)$ and $H_{c2}==\phi_0/2\pi\xi^2$, both of which can be combined to write $H_{c1}/H_{c2}=\ln\kappa/2\kappa^2$, where the symbols have their usual meanings. From the values of $H_{c1}$ and $H_{c2}$ obtained in the present measurements, a value of 26 is obtained for $\kappa$ which is the same as elsewhere also [5].

Figure 5 shows magnetization hysteresis loops obtained at various temperatures using a squid magnetometer. The results obtained using a vibrating sample magnetometer are quite similar and are not shown. The super-current density, according to Bean's critical state model, is given by $J_c=30\times\Delta M/d$, where $\Delta M$ is the hysteresis of magnetization per unit volume (emu/cm$^3$) at a given field and *d* is the mean size of the particles. Assuming an average particle size of $5\times10^{-3}$ cm, the critical current densities obtained at various fields and temperatures are plotted in Fig. 6. It is seen that the critical current densities are fairly large even in 1 Tesla applied field and are of the same order of magnitude as reported elsewhere [9].

**Figure Captions**

1. Field-cooled and zero-field-cooled magnetization of $MgB_2$ as a function of temperature in an applied field of 50 Oe showing the superconducting transition temperature.

2. Low field magnetization of $MgB_2$ as a function of applied field at various temperatures.

3. Plot of the deviation in magnetization, $\Delta M$, (from the low field linear fit, extrapolated to higher fields) versus applied filed for the delineation of $H_{c1}$. Inset shows the values of lower critical field, $H_{c1}$, at various temperatures.

4. Zero-field-cooled magnetization as a function of temperature for $MgB_2$ in various applied fields. Inset shows the plot of upper critical field, $H_{c2}$, as a function of temperature.

5. Magnetization hysteresis loops for $MgB_2$ at various temperatures.

6. Magnetization critical current density versus temperature (in logarithmic scale) for $MgB_2$ at various applied fields.

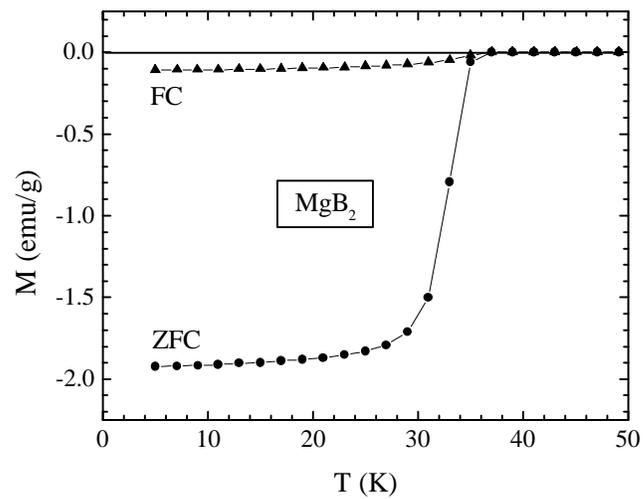

Fig. 1

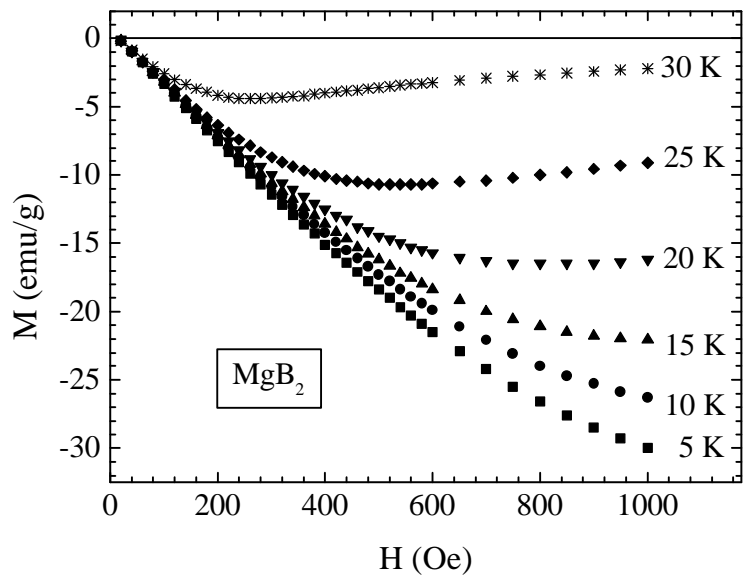

Fig. 2

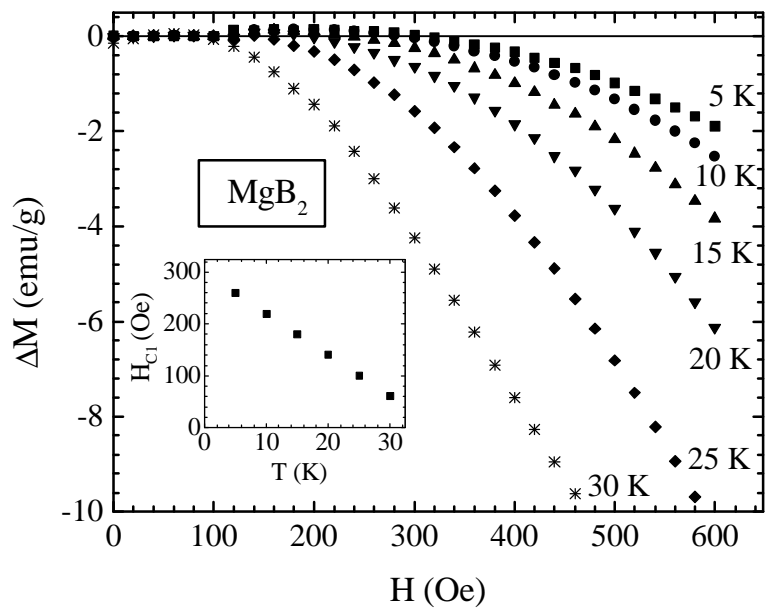

Fig. 3

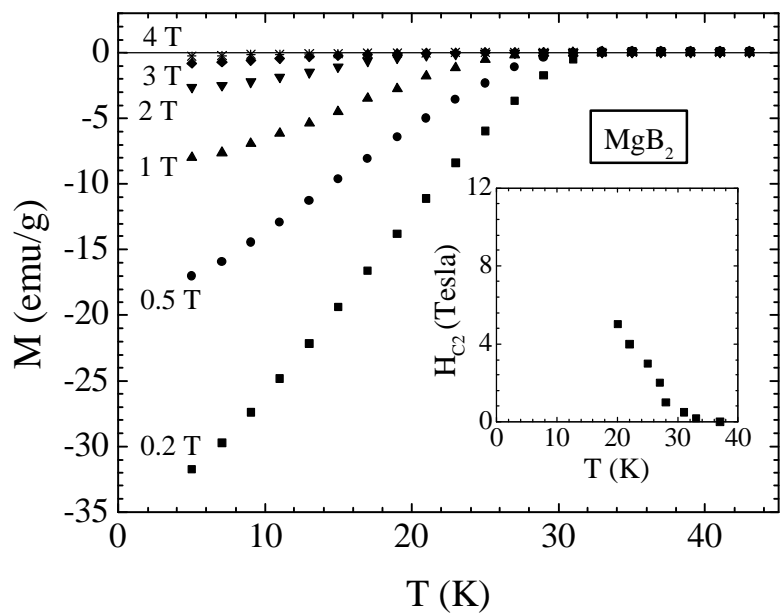

Fig. 4

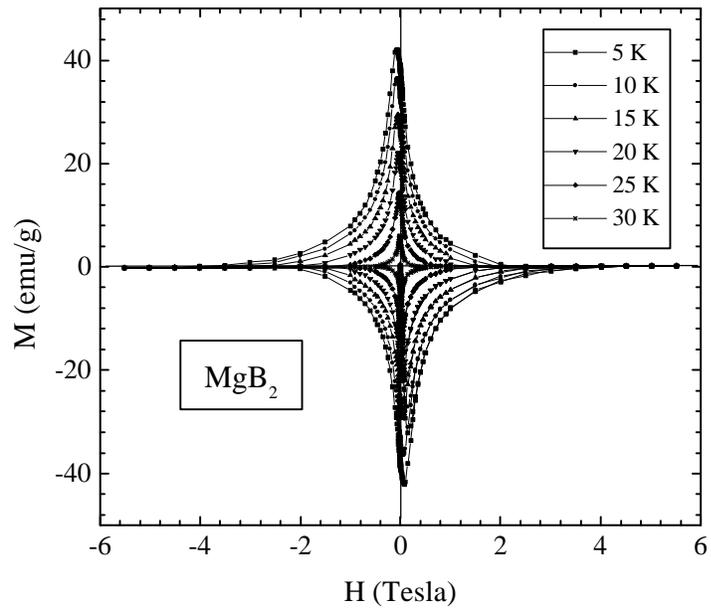

Fig. 5

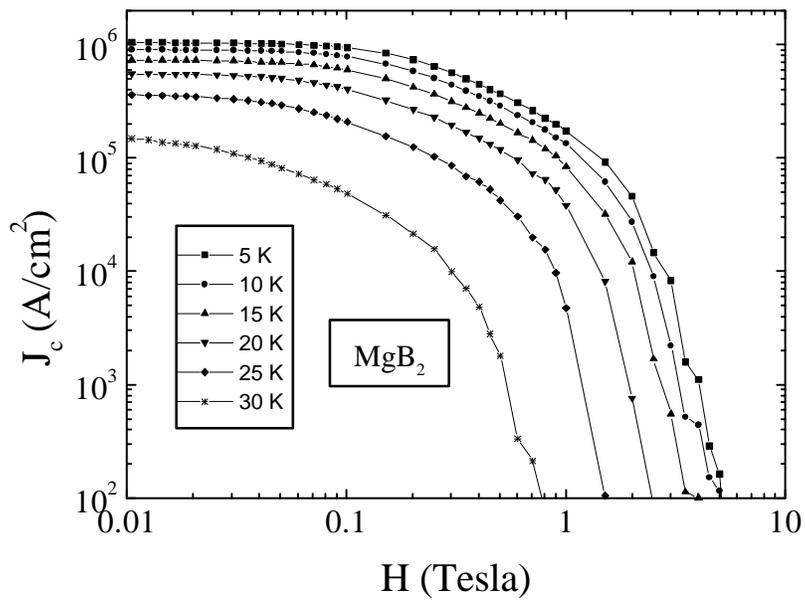

Fig. 6

13